\documentclass{article}

\usepackage{PRIMEarxiv}

\usepackage[T1]{fontenc}    
\usepackage{hyperref}       
\usepackage{url}            
\usepackage{booktabs}       
\usepackage{amsfonts}       
\usepackage{nicefrac}       
\usepackage{microtype}      
\usepackage{lipsum}
\usepackage{fancyhdr}       
\usepackage{graphicx}       
\graphicspath{{media/}}     

\title{Topological data analysis of vortices in the magnetically-induced current density in LiH molecule.
}

\author{
  Ma\l gorzata Olejniczak\\  
  Centre of New Technologies\\
  University of Warsaw\\
  S. Banacha 2c, 02-097 Warsaw, Poland\\
  \texttt{malgorzata.olejniczak@cent.uw.edu.pl} \\
   \And
  Julien Tierny \\
  CNRS, Sorbonne Universit\'e, \\
  Laboratoire d'Informatique de Paris 6, LIP6, \\
  F-75005 Paris, France
}

\begin{document}
\maketitle

\begin{abstract}
A novel strategy for extracting axial (AV) and toroidal (TV) vortices in the magnetically-induced current density (MICD) in molecular systems is introduced, and its pilot application to LiH molecule is demonstrated. It exploits differences in the topologies of AV and TV cores and involves two key steps: selecting a scalar function that can describe vortex cores in MICD and its subsequent topological analysis. The scalar function of choice is  $\Omega^{B_\alpha}$ based on the velocity-gradient $\Omega$ method known in research on classical flows. The Topological Data Analysis (TDA) is then used to analyze the $\Omega^{B_\alpha}$ scalar field. In particular, TDA robustly assigns distinct topological features of this field to different vortex types in LiH: AV to saddle-maximum separatrices which connect maxima to 2-saddles located on the domain's boundary, TV to a  1-cycle of the super-level sets of the input data. Both are extracted as the most \emph{persistent} features of the topologically-simplified $\Omega^{B_\alpha}$ scalar field.
\end{abstract}


\section{Introduction}

Magnetic perturbations induce a rotational motion of electrons in molecules, and the strength, shape, and spatial extent of emerging vortical structures are unique molecular fingerprints. The analysis of molecular response to such perturbations is valuable for understanding molecular magnetic properties and in studies on electron delocalization, aromaticity, inter- and intra-atomic interactions, molecular structure, or reaction mechanisms.\cite{sundholm-wcms-6-639-2016,sundholm-cc-57-12362-2021} The magnetically-induced electron current density (MICD) is the primary quantum mechanical property to consider in this analysis. It is composed of induced vortices whose investigation in real space grew to be a large part of the Quantum Chemical Topology (QCT).\cite{leszczynski-topochem-book-2016} 

This work focuses on a particular case of an external, static, weak, and homogeneous magnetic field perturbation. In this setting, MICD is a non-symmetric second-order tensor, defined as a first-order derivative of the electron current density, $\vec{J}$, with respect to the magnetic field, $\vec{B}$, at the zero-field limit; $J^B = [J_\beta^{B_\alpha}]_{\alpha\beta\in\{x,y,z\}} = \left. \frac{d J_\beta}{dB_\alpha} \right|_{\vec{B} = 0}$. The analysis of such tensors is technically and conceptually demanding,\cite{hergl-cgf-40-135-2021} which motivates the search for simpler descriptors of MICD represented by vector or scalar fields.

\begin{figure*}
\centering
\includegraphics[width=\linewidth]{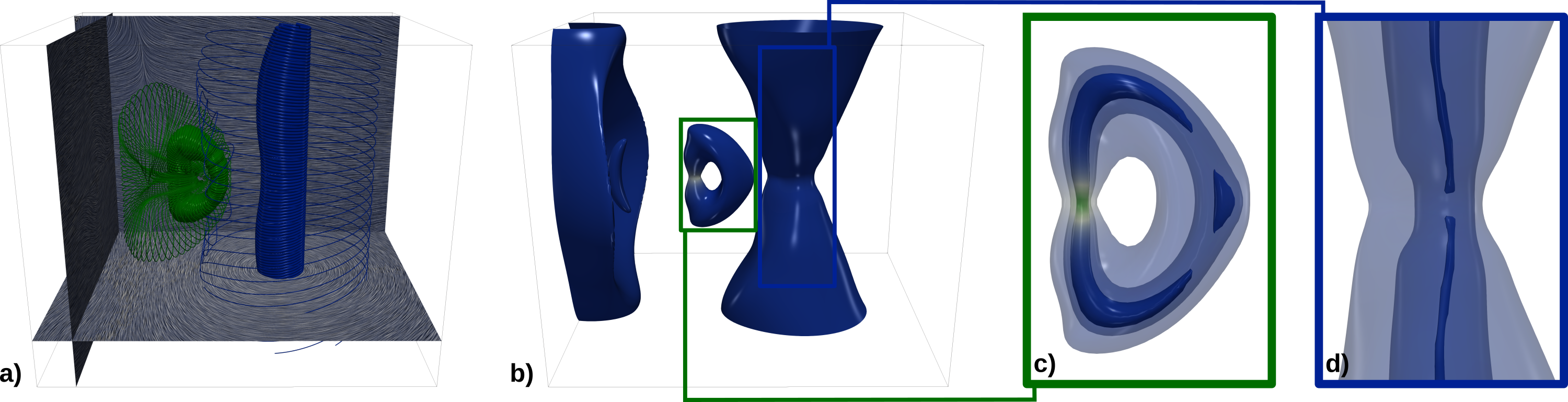}
\caption{
Vortex analysis with $\Omega^{B_\perp}$.
(a) The input vector field $\vec{J}^{B_\perp}$ (shown with the line integral convolution, planes at the bottom, back and left side) includes two vortices: an axial vortex (highlighted by two sets of periodic orbits, in blue) and a toroidal vortex (highlighted by two sets of periodic orbits, in green). (b) The scalar field $\Omega^{B_\perp}$ defines isosurfaces (at isovalue $0.5$, here colored by 
$\perp$-component of the curl of $\vec{J}^{B_\perp}$ field
)
which capture well the geometry of the vortices.
However, the precise extraction of the vortex center lines is not possible with the isosurfaces (c, d) since these disconnect in the vicinity of the vortex cores.}
\label{fig_input}
\end{figure*}

A popular approach\cite{pelloni-pra-74-012506-2006} is to select the perturbation direction, $\alpha$, and consider the $\vec{J}^{B_\alpha}$ vector. It is particularly informative for symmetric systems where $\alpha$ can be fixed {as perpendicular} to a plane aligned with a meaningful molecular feature, such as an aromatic ring or a chemical bond.\footnote{We use $J^B$ and $\vec{J}^{B_\alpha}$ symbols to refer to a tensor and a vector quantity, respectively; the superscript ``${B_\alpha}$'' emphasizes the dependency on a fixed perturbation direction.} Nevertheless, either direct or topological, such analysis can still be demanding, and its complexity increases for larger and non-symmetric molecular systems.\cite{sundholm-wcms-6-639-2016,sundholm-cc-57-12362-2021} 
Reduction of $J^B$ or $\vec{J}^{B_\alpha}$ to scalar functions entails a further loss of information. Here, focusing on the description of a particular vortex feature may be helpful, for instance, (i) the location of a vortex center, (ii) the shape of a vortex core, (iii) and of the boundary between vortices (``separatrix''), (iv) the region of influence of a vortex (``basin''), (v) the rotational axis and circulation direction around that axis, or (vi) the absolute and relative strength of a vortex. Specifying the feature of interest helps to choose a method for vortex extraction. 

The goal of this work is a qualitative description of vortices through the topologies of vortex cores, for which we report the following new contributions:
\begin{itemize}
\item \emph{Proposal of a new scalar function to analyze MICD in molecular systems.} Motivated by the interpretation of $\vec{J}$ as a velocity density,\cite{sundholm-cc-57-12362-2021}{\cite{lazzeretti-pnmrs-36-1-2000}} and by advances in vortex tracking techniques, we derive the $\Omega^{B_\alpha}$ scalar function from the $\vec{J}^{B_\alpha}$ vector field. It is analogous to the $\Omega$ index used in research on classical flows \cite{liu-scpma-59-684711-2016,dong-jh-30-541-2018} and, to our best knowledge, has not been considered in MICD studies.
\item \emph{Topological Data Analysis of the $\Omega^{B_\alpha}$ scalar function.} Motivated by the success of TDA to describe scalar data through its topological features in numerous applications\cite{bremer_tvcg11, Lukasczyk17, soler_ldav19, chemistry_vis14, Malgorzata19, sousbie11}
including vortex extraction,\cite{Bridel-Bertomeu19, nauleau_ldav22} we apply TDA tools to the $\Omega^{B_\alpha}$ scalar field. To our best knowledge, it is the first report on employing TDA in the MICD context.
\end{itemize}

\section{Results and discussion}
\label{sec:results}

Lithium hydride, LiH, is a classic test system for the analysis of MICD. The current density induced by a magnetic field { perpendicular} to the Li-H bond (from now on marked as ${B_\perp}$) was examined already in 1960s,\cite{stevens-jcp-40-2238-1964} and has been a subject of numerous analyses ever since.\cite{keith-jcp-99-3669-1993,pelloni-tca-123-353-2009,summa-jcp-156-154105-2022,berger-pccp-24-23089-2022} 

A lot can be learned from the $\vec{J}^{B_\perp}$ vector field's topology.  
The eigenvalues of the gradient of $\vec{J}^{B_\perp}$, represented by a Jacobian matrix, $\nabla \vec{J}^{B_\perp} = \left[\partial_\gamma J_\beta^{B_\perp}\right]_{\perp;\beta\gamma}$, where $\beta,\gamma\in\{x,y,z\}$, and $\partial_\gamma$ is a shorthand notation for spatial derivatives, inform about possible types of critical points (CPs) of $\vec{J}^{B_\perp}$ field.\cite{lazzeretti-pnmrs-36-1-2000} They can be isolated or form open or closed paths - in QCT referred to as ``stagnation lines (SLs)''.\cite{gomes-jcp-78-4585-1983,lazzeretti-rl-30-515-2019} Open SLs extending to the domain's boundaries act as centers of axial vortices, while closed SLs forming loops lead to toroidal vortices. Isolated CPs and SLs form a topological skeleton, the so-called ``stagnation graph (SG)''.\cite{gomes-pra-28-559-1983} { The direction of rotation of the induced flow on the planes perpendicular to SLs, formally defined as a local curl of the $\vec{J}^{B_\perp}$, can be clockwise or counterclockwise for the observer toward whom the $B_\perp$ points; such a flow is referred to as "diatropic" or "paratropic", respectively.\cite{monaco-pccp-18-11800-2016, keith-jcp-99-3669-1993}} 

{ The} SG of $\vec{J}^{B_\perp}$ in LiH shows two distinct non-overlapping vortices: AV established by the diatropic circulation in the H atomic basin and extending to the whole domain and TV composing of the diatropic and paratropic currents induced in the Li atomic basin. The centers of these two vortices are represented by SLs of different topologies: for AV the SL is a straight line, and for TV - the SL is a closed loop perpendicular to the Li-H bond.\cite{pelloni-tca-123-353-2009,berger-pccp-24-23089-2022}

\begin{figure*}[t]
\centering
\includegraphics[width=\linewidth]{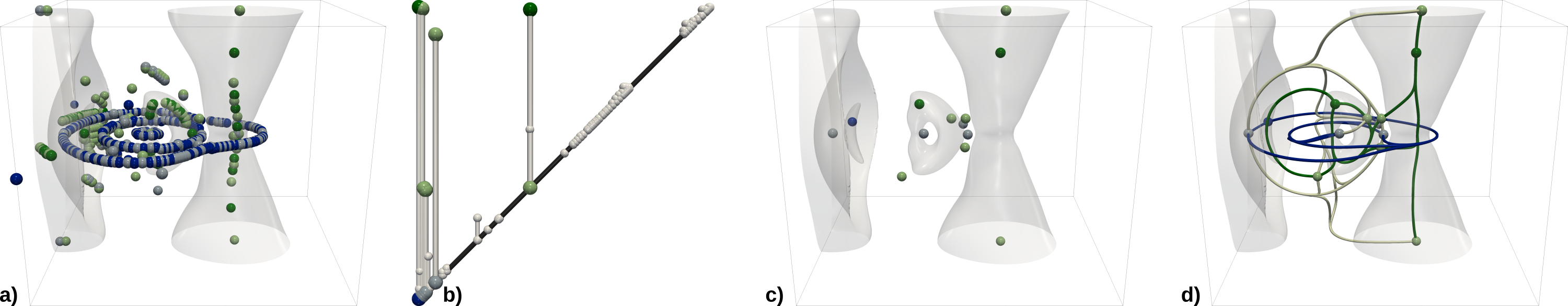}
\caption{Topological analysis of the scalar field $\Omega^{B_\perp}$. (a) Initially, $\Omega^{B_\perp}$ counts many critical points (dark blue: minima, light blue: $1$-saddles, light green: $2$-saddles, dark green: maxima), which are mostly due to sampling artifacts.
(b) The persistence diagram displays the critical points in pairs (vertical bars) responsible for the creation and destruction of a single topological feature (e.g. connected component, handle, void) of the sub-level sets of $\Omega^{B_\perp}.$ The $X$-value of a bar indicates the scalar value of creation of the corresponding feature (usually called its \emph{"birth"}), while its top $Y$-value indicates the scalar value of its destruction (usually called its \emph{"death"}). The height of each bar is the difference in $\Omega^{B_\perp}$ values between its two critical points. It is an established measure of importance called \emph{Topological Persistence}. Large bars correspond to salient features while small bars 
(
with a persistence smaller than $30 \%$ of the function range, shown with white smaller tubes and spheres) correspond to noise. These can be easily discarded from the data. (c) The input data is topologically-simplified by removing the noisy features identified in the diagram (white spheres in (b)), leading to only a few \emph{meaningful} critical points. (d) The one-dimensional separatrices of the Morse-Smale complex (blue: minimum-saddle, white: saddle-saddle, green: saddle-maximum) are computed for the topologically-simplified data. These curves capture \emph{extremal lines} within the data, located at the center of isosurfaces. In particular, a subset of the saddle-maximum separatrices (green) seem to capture the center lines of the vortices.}
\label{fig_topology}
\end{figure*}

With this picture emerging from the topology of $\vec{J}^{B_\perp}$, we now explore the $\Omega^{B_\perp}$ scalar function. Its derivation is inspired by one of the velocity-gradient-based vortex extraction techniques.\cite{liu-scpma-59-684711-2016,dong-jh-30-541-2018} Such methods exploit the decomposition of the gradient of the velocity field into its symmetric and antisymmetric parts and capitalize on two observations: (i) the antisymmetric part of the velocity gradient is related to the vorticity of the velocity field, and (ii) the vorticity describes the rigid-body circulation \textit{and} shear.\cite{liutex_vortex_book_2021}
{ The} omega index ($\Omega^{B_\perp}$), is defined as the ratio of the {square of } the norm of the antisymmetric part of the velocity gradient tensor ($\nabla \vec{J}^{\;A;B_\perp}$) to the sum of {squares of } norms of symmetric ($\nabla \vec{J}^{\;S;B_\perp}$) and antisymmetric ($\nabla \vec{J}^{\;A;B_\perp}$) parts of this tensor; Frobenius norms, $||\cdot||_F$, are employed,
\begin{equation}
  \Omega^{B_\perp} = \frac{||\nabla \vec{J}^{\;A;B_\perp}||{^2}_F}{||\nabla \vec{J}^{\;A;B_\perp}||{^2}_F + ||\nabla \vec{J}^{\;{S};B_\perp}||{^2}_F}.
  \label{eq:Omega}
\end{equation}
As such, Omega defines a vortex as the area in which vorticity overtakes deformation, and its large values indicate vortex cores.
{ It is worth noting that the numerator of Eq. 1 is related to the anisotropy of the antisymmetric part of the Jacobian of the $\vec{J}^{B_\alpha}$ vector, which is one of the second-rank tensor invariants. Using tensor invariants to define scalar descriptors for analyzing the magnetically-induced current density is of increasing interest;\cite{lazzeretti-jcp-148-134109-2018} examples of tensor's anisotropy measures employed in MICD analysis involve the popular ACID index\cite{herges-jpca-105-3214-2001,geuenich-cr-105-3758-2005} and the recently proposed AACID function.\cite{monaco-jpca-122-4681-2018,monaco-pccp-21-11564-2019}}

To obtain the $\Omega^{B_\perp}$ scalar function for LiH, we applied \autoref{eq:Omega} to the magnetically-induced current density and its gradient calculated in the DIRAC\cite{DIRAC22,saue-jcp-152-204104-2020-diracpaper} software (details in ESI).
We first explore the ability of $\Omega^{B_\perp}$ to represent vortex cores in LiH by plotting isosurfaces of this field in
\autoref{fig_input}(b).
A sample isosurface corresponding to $\Omega^{B_\perp}=0.5$ 
indeed follows the pattern of SLs in { the} $\vec{J}^{B_\perp}$ field, exhibiting a pea-pod shape in the H atom basin and a circular shape in the Li atom basin. 
However, 
as shown in \autoref{fig_input}, while the shape of the isosurfaces of $\Omega^{B_\perp}$ capture well the geometry of the vortices, the precise extraction of the exact geometry of the vortex center lines is not possible with isosurfaces, since these disconnect in the vicinity of the vortex cores. 

To address this issue, we employ tools from Topological Data Analysis, which is a recent area of research at the interface between mathematics and computer science \cite{edelsbrunner09}. It provides a unified, robust and multi-scale set of tools for the extraction of structural patterns in data. 
In particular, for the following experiments, we used the \emph{Topology ToolKit} (TTK)  \cite{ttk17, ttk19}, which is an open-source package for Topological Data Analysis (see \url{https://topology-tool-kit.github.io/}).

First,
we  employ Discrete Morse Theory (DMT) \cite{forman98, robins_pami11} to extract the critical points of $\Omega^{B_\perp}$ (\autoref{fig_topology}(a)). 
In contrast to numerical approaches, which can be prone to numerical instabilities, DMT is a combinatorial formalism which robustly identifies the critical points of the 
scalar field
as cells of the 
input 
domain (each critical point of index $i$ will be extracted on an $i$-dimensional cell). Then, local minima will be identified on vertices, $1$-saddles on edges, $2$-saddles on faces, etc. 
For example, a vertex of the domain with a lower scalar value than all its neighbor vertices will be identified by DMT as a local minimum.
We refer the reader to specific references\cite{forman98, robins_pami11} for further technical details, generalizations and algorithms.

However, this
first raw analysis shows a large number of critical points (\autoref{fig_topology}(a)).
These are mostly due to discretization artifacts related to the sampling of a scalar field defined as a function of norms (\autoref{eq:Omega}) along the fixed, three-dimensional regular grid modeling the input domain.  
When $||\nabla \vec{J}^{\;A;B_\perp}||$ vanishes,  $\Omega^{B_\perp}$ tends to zero. 
In the vicinity of these points, some vertices of the input regular grid may have a value of $\Omega^{B_\perp}$ very close to (but greater than) zero and it is possible that these vertices have a lower $\Omega^{B_\perp}$-value than all their neighbors. Such vertices will be identified by DMT as local minima (dark blue spheres). Since DMT is globally consistent  (i.e. the alternating sum of numbers of critical points is equal to the Euler characteristic of the grid),  each of these spurious minima will be accompanied by a corresponding spurious $1$-saddle (light blue spheres). Similar observations can be made for pairs of $1$ and $2$-saddles (close, spurious light blue and light green spheres) or pairs of $2$-saddles and maxima (close, spurious light green and dark green spheres).

To address this sampling noise, we use in a second stage Persistent Homology. 
In particular,
we compute the persistence diagram \cite{edelsbrunner09}
(\autoref{fig_topology}(b)), which helps identify the critical points which are the most \emph{meaningful} (i.e. persistent, or salient). 
Specifically, the persistence diagram pairs the critical point $c$ of index $i$ responsible for the creation of an $i$-dimensional cycle in the sublevel sets of the input scalar field, with the critical point $c'$ of index $i+1$ responsible for the filling of that cycle. In the diagram, each pair of critical point $(c, c')$ of index $i$ and $i+1$ is represented as a vertical bar, whose extremities are embedded at locations $\big(\Omega^{B_\perp}(c), \Omega^{B_\perp}(c)\big)$ and $\big(\Omega^{B_\perp}(c), \Omega^{B_\perp}(c')\big)$. The height of the bar, $\Omega^{B_\perp}(c') - \Omega^{B_\perp}(c)$, is an established measure of importance of the critical point pair $(c, c')$ called \emph{Topological Persistence}. We refer the interested reader to a reference textbook\cite{edelsbrunner09} on the fundamentals of Persistent Homology and to a specific reference\cite{guillou22} for 
an efficient computation algorithm (which we used in our experiments).

Next, a topologically-simplified version of $\Omega^{B_\perp}$ is computed \cite{Lukasczyk_vis20} (\autoref{fig_topology}(c)), to remove 
all the critical point pairs of low persistence (corresponding to noise). This results in a smoother field, which has far less critical points (\autoref{fig_topology}(c)).
%

Last, from
this topologically-simplified version of $\Omega^{B_\perp}$, the $1$-dimensional separatrices of the Morse-Smale complex (MSC) are computed (\autoref{fig_topology}(d)). 
In short, the MSC is a cellular decomposition of the domain
such that, for all the points of a given cell of the MSC, the forward and backward integrations of the field's gradient terminate at the same pair of critical points (see \cite{ttk17} for formal definitions). In this complex, $1$-dimensional cells (called separatrices) are curves which are tangential to the gradient, and which connect a critical point of index $i$ to a critical point of index $i+1$. In the applications, these curves often capture "extremal lines" in the data (i.e. curves which are locally minimal/maximal within an intersection plane orthogonal to the curve's tangent). These curves are often used to extract the center lines of isosurfaces which have a tubular shape \cite{sousbie11} (such as the isosurfaces of $\Omega^{B_\perp}$).
Specifically, the saddle-maximum separatrices (green  integral lines linking $2$-saddles to maxima) seem to already capture the geometry of the vortices.

\begin{figure}
\centering
\includegraphics[width=\linewidth]{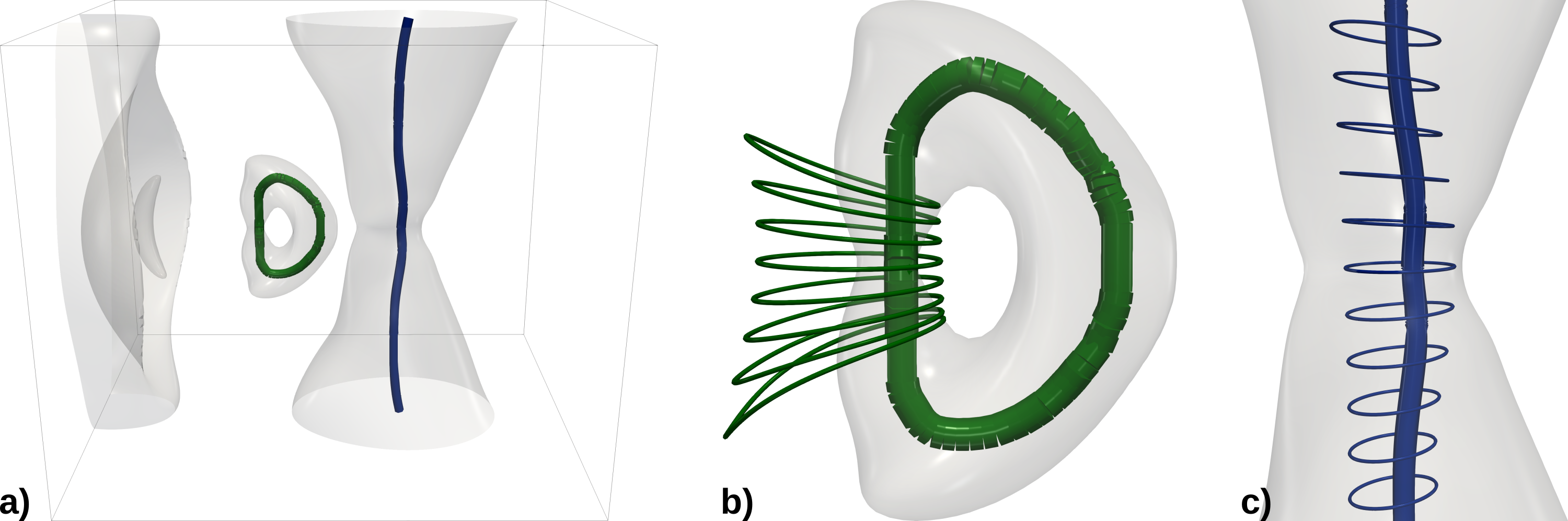}
\caption{Vortex extraction with the topological analysis of $\Omega^{B_\perp}$. 
(a) 
First, the axial vortex is automatically extracted (blue curve) as  the saddle-maximum separatrices of the topologically-simplified data which connect a maximum to $2$-saddles located on the boundary of the grid.
Second, the toroidal vortex is automatically extracted (green curve) as the most persistent $1$-cycle in the super-level sets of $\Omega^{B_\perp}$ (which is a generator of the topological handle of the solid torus shown with the grey isosurface).
These curves capture precisely the center lines of the vortices, as illustrated in (b) and (c) with a set of periodic orbits of $\Omega^{B_\perp}$ (thin curves).}
\label{fig_output}
\end{figure}

We refine this analysis in \autoref{fig_output}. First, we make the hypothesis that the length of an axial vortex is much larger than the scale of the three-dimensional regular grid forming the domain of $\Omega^{B_\perp}$. Thus, an axial vortex will necessarily interesect the boundary of the grid. Hence, we automatically extract the axial vortex by collecting the separatrices linking boundary $2$-saddles to maxima (blue curve, \autoref{fig_output}(a)). Next, we make the hypothesis that a toroidal vortex will necessarily correspond to solid torus in the super-level sets of $\Omega^{B_\perp}$ (c.f. grey isosurfaces). Hence, we automatically extract the toroidal vortex by collecting the most persistent generator of the $1$-dimensional persistent homology group of the opposite of $\Omega^{B_\perp}$ \cite{guillou22} (green curve, \autoref{fig_output}(a)). These curves capture precisely the center lines of the vortices, as illustrated in \autoref{fig_output}(b) and \autoref{fig_output}(c), where a set of periodic orbits of $\Omega^{B_\perp}$ (thin curves) is exactly located around them.

We base our hypotheses on the shapes of SLs of the $\vec{J}^{B_\perp}$ vector field, which guide the selection of TDA abstractions representing the information on vortex cores encoded in the $\Omega^{B_\perp}$ scalar field. As a result, vortices extracted from the topology of the $\Omega^{B_\perp}$ scalar field in LiH are consistent with expectations based on a more complex vector field analysis.
This protocol should be further confirmed and adapted to larger and non-symmetric molecules in which the $\vec{J}^{B_\perp}$ vector field exhibits a disconnected SG.

In such cases, the induced vortices can also be associated with the isolated CPs of the $\vec{J}^{B_\alpha}$ vector field and involve the flow that spirals about the rotation axis.\cite{pelloni-ijqc-111-356-2011} Further extension of this work is to consider other scalar functions (e.g., independent of the fixed perturbation direction) successfully used to track vortex cores.\cite{Gunther2018} We also preview the application of TDA to functions representing other features of induced vortices, e.g., MICD basins, which are of great interest in QCT.{\cite{monaco-jpca-123-1558-2019}} The quantitative description of induced vortices through numerous integration schemes or TDA-specific measures is also invaluable.

\section{Conclusions}
This work presented exploratory studies on extracting vortices in the magnetically-induced current density in LiH molecule. It proposed a novel procedure involving (i) selecting a scalar field derived from the MICD tensor and (ii) applying the Topological Data Analysis tools to this field.

We presented the application of this protocol to the current density induced in LiH molecule by a magnetic field perpendicular to the chemical bond. It exhibits only two vortices (one axial and one toroidal), yet, it allows us to make a few observations on the pros and cons of our strategy.

Its main strength is the use of TDA. We demonstrated which TDA abstractions can extract vortices whose centers are straight lines (AV) or closed loops (TV). The extension of the TDA protocol to describe vortices in MICD of more complex systems remains to be verified; it may involve other scalar functions better suited to track vortex centers irrespective of their types. The ability of TDA to describe data features at various scales \cite{edelsbrunner09} is another factor promising for complex systems, which may exhibit many vortices of varying strengths and spatial distributions. 
Due to the combinatorial nature of TDA algorithms, our approach is efficient, consistent and multi-scale (i.e. in terms of the importance of the critical points).
Also, in contrast to numerical schemes currently applied to analyze magnetically-induced currents (often based on Runge-Kutta methods),\cite{lazzeretti-pnmrs-36-1-2000} TDA is free from numerical instabilities given its combinatorial nature.

Note, also,
that TDA does not involve any empirical parameters, such as the one introduced initially in the definition of the $\Omega$ index.\cite{liu-scpma-59-684711-2016}

Another strength of our strategy is that it provides a compact description of induced vortices without the need to analyze the precedent $J^B$ or $\vec{J}^{B_\alpha}$ fields. Instead, we applied TDA to a scalar function derived from these fields. Working on scalar fields has many advantages, yet it may raise concerns about the information lost in reducing higher-rank tensor fields
to scalar fields. In some cases, this information can be reintroduced to final visualizations, e.g., as colors of isosurfaces (which could represent the direction of circulation of induced flows, here omitted for clarity).

Focusing on a particular feature of vortices in MICD can help select an adequate scalar function; in this task, other research domains studying rotational motion can be immensely inspiring. In our protocol, we adapted the $\Omega$ index used in studies on classical flows. { The main drawback of the resultant $\Omega^{B_\alpha}$ function is the increased computational time and memory due to the need to compute the $\vec{J}^{B_\alpha}$ gradient (however, this quantity is computed in other topological analyses of the $\vec{J}^{B_\alpha}$ vector field as well). Also, the dependency of the $\Omega^{B_\alpha}$ on the fixed perturbation direction (we derived $\Omega^{B_\alpha}$ from $\vec{J}^{B_\alpha}$ for a fixed $\alpha$) can be limiting.}
While this function served our purpose of describing vortex cores, addressing other features may require different functions. For instance, extracting basins of vortices needs functions that well-represent flow far from vortex cores. We can, however, use the approach to focus on a single feature of MICD to our advantage. For instance, we expect $\Omega^{B_\alpha}$ to be a good indicator of \emph{localized} flow domains, worth exploring in the context of localized/semi-localized/delocalized currents related to chemical bonds or molecular fragments.

\section*{Data availability statement}
The details on the quantum chemistry calculations, the analysis protocol, and the data reported in this work are available { on the dedicated website (\href{https://tda-qchem.github.io/tda-qchem-examples/LiH_MICD/}{`https://tda-qchem.github.io/tda-qchem-examples/LiH\_MICD/`} and on} zenodo (\href{https://doi.org/10.5281/zenodo.7446735}{DOI:10.5281/zenodo.7446735}.)

\section*{Author Contributions}
MO: 
conceptualization, 
data curation, 
funding acquisition, 
investigation, 
methodology, 
software, 
validation, 
writing – original draft, 
writing – review \& editing
; 
JT: 
conceptualization, 
formal analysis, 
funding acquisition, 
investigation, 
methodology, 
software, 
validation, 
visualization, 
writing – review \& editing
(according to \href{https://groups.niso.org/higherlogic/ws/public/download/26466/ANSI-NISO-Z39.104-2022.pdf}{ANSI/NISO Z39.104-2022}.)

\section*{Conflicts of interest}
There are no conflicts to declare.

\section*{Acknowledgements}
{We thank the referees for carefully reading our manuscript and providing insightful comments that helped us substantially improve the quality of this paper.}
We acknowledge support from the Polish National Science Centre (NCN) (grant 
number 2020/38/E/ST4/00614) and the European Commission grant 
ERC-2019-COG 
\emph{``TORI''} (ref. 863464, 
\url{https://erc-tori.github.io/}).
{ This research was supported in part by PLGrid Infrastructure.}

\bibliographystyle{unsrt}  
\bibliography{manuscript_arxiv}

\begin{thebibliography}{10}

\bibitem{sundholm-wcms-6-639-2016}
Dage Sundholm, Heike Fliegl, and Raphael J.F.~F Berger.
\newblock {Calculations of magnetically induced current densities: theory and
  applications}.
\newblock {\em Wiley Interdiscip Rev Comput Mol Sci}, 6(6):639--678, nov 2016.

\bibitem{sundholm-cc-57-12362-2021}
Dage Sundholm, Maria Dimitrova, and Raphael~J.F. Berger.
\newblock Current density and molecular magnetic properties.
\newblock {\em ChemComm}, 57:12362--12378, 11 2021.

\bibitem{leszczynski-topochem-book-2016}
Jerzy Leszczynski.
\newblock {\em Applications of Topological Methods in Molecular Chemistry}.
\newblock Springer International Publishing, 2016.

\bibitem{hergl-cgf-40-135-2021}
Chiara Hergl, Christian Blecha, Vanessa Kretzschmar, Felix Raith, Fabian
  G\"unther, Markus Stommel, Jochen Jankowai, Ingrid Hotz, Thomas Nagel, and
  Gerik Scheuermann.
\newblock Visualization of tensor fields in mechanics.
\newblock {\em Comput Graph Forum}, 40:135--161, 9 2021.

\bibitem{pelloni-pra-74-012506-2006}
Stefano Pelloni, Francesco Faglioni, Riccardo Zanasi, and Paolo Lazzeretti.
\newblock {Topology of magnetic-field-induced current-density field in
  diatropic monocyclic molecules}.
\newblock {\em Phys. Rev. A}, 74(1):012506, jul 2006.

\bibitem{lazzeretti-pnmrs-36-1-2000}
P.~Lazzeretti.
\newblock {Ring currents}.
\newblock {\em Prog. Nucl. Mag. Res. Sp.}, 36(1):1--88, 2000.

\bibitem{liu-scpma-59-684711-2016}
Chao~Qun Liu, Yi~Qian Wang, Yong Yang, and Zhi~Wei Duan.
\newblock New omega vortex identification method.
\newblock {\em {Sci. China: Phys. Mech. Astron.}}, 59:684711, 6 2016.

\bibitem{dong-jh-30-541-2018}
Xiang-rui Dong, Yi-qian Wang, Xiao-ping Chen, Yinlin Dong, Yu-ning Zhang, and
  Chaoqun Liu.
\newblock Determination of epsilon for omega vortex identification method.
\newblock {\em J Hydrodyn}, 30:541--548, 7 2018.

\bibitem{bremer_tvcg11}
P.T. Bremer, G.~Weber, J.~Tierny, V.~Pascucci, M.~Day, and J.~Bell.
\newblock Interactive exploration and analysis of large scale simulations using
  topology-based data segmentation.
\newblock {\em TVCG}, pages 1307--1324, 2011.

\bibitem{Lukasczyk17}
Jonas Lukasczyk, Garrett Aldrich, Michael Steptoe, Guillaume Favelier, Charles
  Gueunet, Julien Tierny, Ross Maciejewski, Bernd Hamann, and Heike Leitte.
\newblock Viscous fingering: A topological visual analytic approach.
\newblock In {\em PMVMSP}, pages 9--19, 2017.

\bibitem{soler_ldav19}
Maxime Soler, Martin Petitfrere, Gilles Darche, Melanie Plainchault, Bruno
  Conche, and Julien Tierny.
\newblock {Ranking Viscous Finger Simulations to an Acquired Ground Truth with
  Topology-Aware Matchings}.
\newblock In {\em LDAV}, 2019.

\bibitem{chemistry_vis14}
D.~Guenther, R.~Alvarez-Boto, J.~Contreras-Garcia, J.-P. Piquemal, and
  J.~Tierny.
\newblock Characterizing molecular interactions in chemical systems.
\newblock {\em TVCG}, pages 2476--2485, 2014.

\bibitem{Malgorzata19}
Malgorzata Olejniczak, Andr\'e Severo~Pereira Gomes, and Julien Tierny.
\newblock {A Topological Data Analysis Perspective on Non-Covalent Interactions
  in Relativistic Calculations}.
\newblock {\em Int. J. Quant. Chem.}, 120:e26133, 2019.

\bibitem{sousbie11}
T.~Sousbie.
\newblock The persistent cosmic web and its filamentary structure: Theory and
  implementations.
\newblock {\em MNRAS}, 414:1--38, 2011.

\bibitem{Bridel-Bertomeu19}
Thibault Bridel{-}Bertomeu, Benjamin Fovet, Julien Tierny, and Fabien
  Vivodtzev.
\newblock {Topological Analysis of High Velocity Turbulent Flow}.
\newblock In {\em LDAV}, 2019.

\bibitem{nauleau_ldav22}
Florent Nauleau, Fabien Vivodtzev, Thibault Bridel{-}Bertomeu,
  H{\'{e}}lo{\"{\i}}se Beaugendre, and Julien Tierny.
\newblock {Topological Analysis of Ensembles of Hydrodynamic Turbulent Flows -
  An Experimental Study}.
\newblock In {\em LDAV}, 2022.

\bibitem{stevens-jcp-40-2238-1964}
Richard~M. Stevens and William~N. Lipscomb.
\newblock {Perturbed Hartree—Fock Calculations. II. Further Results for
  Diatomic Lithium Hydride}.
\newblock {\em J. Chem. Phys.}, 40(8):2238--2247, apr 1964.

\bibitem{keith-jcp-99-3669-1993}
Todd~A. Keith and Richard F.~W. Bader.
\newblock {Topological analysis of magnetically induced molecular current
  distributions}.
\newblock {\em J. Chem. Phys.}, 99(5):3669--3682, sep 1993.

\bibitem{pelloni-tca-123-353-2009}
Stefano Pelloni, Paolo Lazzeretti, and Riccardo Zanasi.
\newblock {Topological models of magnetic field induced current density field
  in small molecules}.
\newblock {\em Theor. Chem. Acc.}, 123(3-4):353--364, feb 2009.

\bibitem{summa-jcp-156-154105-2022}
Francesco~Ferdinando Summa, Guglielmo Monaco, Riccardo Zanasi, and Paolo
  Lazzeretti.
\newblock Origin independent current density vector fields induced by
  time-dependent magnetic field. i. the lih molecule.
\newblock {\em J. Chem. Phys.}, 156:154105, 3 2022.

\bibitem{berger-pccp-24-23089-2022}
Raphael J~F Berger and Maria Dimitrova.
\newblock A natural scheme for the quantitative analysis of the magnetically
  induced molecular current density using an oriented flux-weighted stagnation
  graph. i. a minimal example for lih.
\newblock {\em Phys. Chem. Chem. Phys.}, 24:23089, 2022.

\bibitem{gomes-jcp-78-4585-1983}
J.~A. N.~F. Gomes.
\newblock {Topological elements of the magnetically induced orbital current
  densities}.
\newblock {\em J. Chem. Phys.}, 78(7):4585--4591, apr 1983.

\bibitem{lazzeretti-rl-30-515-2019}
Paolo Lazzeretti.
\newblock {Stagnation graphs and separatrices of local bifurcations in velocity
  and current density planar vector fields}.
\newblock {\em Rendiconti Lincei}, 30(3):515--535, sep 2019.

\bibitem{gomes-pra-28-559-1983}
J.~A. N.~F. Gomes.
\newblock {Topology of the electronic current density in molecules}.
\newblock {\em Phys. Rev. A}, 28(2):559--566, aug 1983.

\bibitem{monaco-pccp-18-11800-2016}
Guglielmo Monaco and Riccardo Zanasi.
\newblock The making of ring currents.
\newblock {\em Phys. Chem. Chem. Phys.}, 18:11800--11812, 2016.

\bibitem{liutex_vortex_book_2021}
Liu Chaoqun and Wang Yiqian, editors.
\newblock {\em Liutex and Third Generation of Vortex Definition and
  Identification}.
\newblock Springer International Publishing, 1 edition, 2021.

\bibitem{lazzeretti-jcp-148-134109-2018}
Paolo Lazzeretti.
\newblock Current density tensors.
\newblock {\em J. Chem. Phys.}, 148:134109, 4 2018.

\bibitem{herges-jpca-105-3214-2001}
Rainer Herges and Daniel Geuenich.
\newblock Delocalization of electrons in molecules †.
\newblock {\em J. Phys. Chem. A}, 105:3214--3220, 2001.

\bibitem{geuenich-cr-105-3758-2005}
Daniel Geuenich, Kirsten Hess, Felix Köhler, and Rainer Herges.
\newblock Anisotropy of the induced current density (acid), a general method to
  quantify and visualize electronic delocalization.
\newblock {\em Chem. Rev.}, 105:3758--3772, 2005.

\bibitem{monaco-jpca-122-4681-2018}
Guglielmo Monaco and Riccardo Zanasi.
\newblock {AACID: Anisotropy of the Asymmetric Magnetically Induced Current
  Density Tensor}.
\newblock {\em J. Phys. Chem. A}, 122:4681--4686, 5 2018.

\bibitem{monaco-pccp-21-11564-2019}
Guglielmo Monaco and Riccardo Zanasi.
\newblock Delocalization energy retrieved from the current density tensor.
\newblock {\em Phys. Chem. Chem. Phys.}, 21:11564--11568, 2019.

\bibitem{DIRAC22}
{DIRAC}, a relativistic ab initio electronic structure program, Release
  {DIRAC22} (2022), written by H.~J.~{\relax Aa}.~Jensen, R.~Bast,
  A.~S.~P.~Gomes, T.~Saue and L.~Visscher, with contributions from I.~A.~Aucar,
  V.~Bakken, C.~Chibueze, J.~Creutzberg, K.~G.~Dyall, S.~Dubillard,
  U.~Ekstr{\"o}m, E.~Eliav, T.~Enevoldsen, E.~Fa{\ss}hauer, T.~Fleig,
  O.~Fossgaard, L.~Halbert, E.~D.~Hedeg{\aa}rd, T.~Helgaker, B.~Helmich--Paris,
  J.~Henriksson, M.~van~Horn, M.~Ilia{\v{s}}, Ch.~R.~Jacob, S.~Knecht,
  S.~Komorovsk{\'y}, O.~Kullie, J.~K.~L{\ae}rdahl, C.~V.~Larsen, Y.~S.~Lee,
  N.~H.~List, H.~S.~Nataraj, M.~K.~Nayak, P.~Norman, G.~Olejniczak, J.~Olsen,
  J.~M.~H.~Olsen, A.~Papadopoulos, Y.~C.~Park, J.~K.~Pedersen, M.~Pernpointner,
  J.~V.~Pototschnig, R.~di~Remigio, M.~Repisky, K.~Ruud, P.~Sa{\l}ek,
  B.~Schimmelpfennig, B.~Senjean, A.~Shee, J.~Sikkema, A.~Sunaga,
  A.~J.~Thorvaldsen, J.~Thyssen, J.~van~Stralen, M.~L.~Vidal, S.~Villaume,
  O.~Visser, T.~Winther, S.~Yamamoto and X.~Yuan (available at
  \url{http://dx.doi.org/10.5281/zenodo.6010450}, see also
  \url{http://www.diracprogram.org}).

\bibitem{saue-jcp-152-204104-2020-diracpaper}
Trond Saue, Radovan Bast, André Severo~Pereira Gomes, Hans Jørgen~Aa. Jensen,
  Lucas Visscher, Ignacio~Agustın Aucar, Roberto~Di Remigio, Kenneth~G. Dyall,
  Ephraim Eliav, Elke Faßhauer, Timo Fleig, Loıc Halbert, Erik~Donovan
  Hedegård, Benjamin Helmich-Paris, Miroslav Iliaš, Christoph~R. Jacob,
  Stefan Knecht, Jon~K. Laerdahl, Marta~L. Vidal, Malaya~K. Nayak, Małgorzata
  Olejniczak, Jógvan Magnus~Haugaard Olsen, Markus Pernpointner, Bruno
  Senjean, Avijit Shee, Ayaki Sunaga, and Joost N.~P. van Stralen.
\newblock The dirac code for relativistic molecular calculations.
\newblock {\em J. Chem. Phys.}, 152:204104, 5 2020.

\bibitem{edelsbrunner09}
H.~Edelsbrunner and J.~Harer.
\newblock {\em Computational Topology: An Introduction}.
\newblock American Mathematical Society, 2009.

\bibitem{ttk17}
Julien Tierny, Guillaume Favelier, Joshua~A. Levine, Charles Gueunet, and
  Michael Michaux.
\newblock The {T}opology {T}ool{K}it.
\newblock {\em TVCG}, pages 832--842, 2017.
\newblock \url{https://topology-tool-kit.github.io/}.

\bibitem{ttk19}
Talha Bin~Masood, Joseph Budin, Martin Falk, Guillaume Favelier, Christoph
  Garth, Charles Gueunet, Pierre Guillou, Lutz Hofmann, Petar Hristov, Adhitya
  Kamakshidasan, Christopher Kappe, Pavol Klacansky, Patrick Laurin, Joshua
  Levine, Jonas Lukasczyk, Daisuke Sakurai, Maxime Soler, Peter Steneteg,
  Julien Tierny, Will Usher, Jules Vidal, and Michal Wozniak.
\newblock {An Overview of the Topology ToolKit}.
\newblock In {\em TopoInVis}, 2019.

\bibitem{forman98}
Robin Forman.
\newblock {A User's Guide to Discrete Morse Theory}.
\newblock {\em Séminaire Lotharingien de Combinatoire}, 48:B48c, 2002.

\bibitem{robins_pami11}
Vanessa Robins, Peter~John Wood, and Adrian~P. Sheppard.
\newblock {Theory and Algorithms for Constructing Discrete Morse Complexes from
  Grayscale Digital Images}.
\newblock {\em {IEEE} Trans. Pattern Anal. Mach. Intell.}, pages 1646--1658,
  2011.

\bibitem{guillou22}
Pierre Guillou, Jules Vidal, and Julien Tierny.
\newblock {Discrete Morse Sandwich: Fast Computation of Persistence Diagrams
  for Scalar Data -- An Algorithm and A Benchmark}.
\newblock {\em ArXiv e-prints}, pages 1--19, 2022.

\bibitem{Lukasczyk_vis20}
Jonas Lukasczyk, Christoph Garth, Ross Maciejewski, and Julien Tierny.
\newblock Localized topological simplification of scalar data.
\newblock {\em TVCG}, pages 572--582, 2020.

\bibitem{pelloni-ijqc-111-356-2011}
Stefano Pelloni and Paolo Lazzeretti.
\newblock Stagnation graphs and topological models of magnetic-field induced
  electron current density for some small molecules in connection with their
  magnetic symmetry.
\newblock {\em Int. J. Quant. Chem.}, 111:356--367, 2 2011.

\bibitem{Gunther2018}
Tobias G{\"{u}}nther and Holger Theisel.
\newblock The state of the art in vortex extraction.
\newblock {\em Computer Graphics Forum}, 37(6):149--173, 2018.

\bibitem{monaco-jpca-123-1558-2019}
Guglielmo Monaco and Riccardo Zanasi.
\newblock Magnetically induced current density spatial domains.
\newblock {\em J. Phys. Chem. A}, 123:1558--1569, 2 2019.

\end{thebibliography}

\end{document}